\documentclass[english,10pt,pra,aps,showpacs,twocolumn,amsmath,amssymb,superscriptaddress,
longbibliography,
floats]{revtex4-2}
\usepackage[T1]{fontenc}
\usepackage[latin9]{inputenc}
\usepackage{verbatim}
\usepackage{refstyle}
\usepackage{amsmath}
\usepackage{amssymb}
\usepackage{graphicx}
\usepackage{color}
\usepackage{bm}
\usepackage{hyperref}
\makeatletter

\usepackage{braket}

\newcommand{\tr}{\mathop{\mathrm{Tr}}}

\usepackage{babel}
\begin{document}

\title{Tan's adiabatic sweep theorem from the variational theorem for the scattering length}

\author{Alexander Yu.~Cherny}
\email{cherny@theor.jinr.ru}
\affiliation{Bogoliubov Laboratory of Theoretical Physics, Joint Institute for Nuclear
Research, 141980, Dubna, Moscow region, Russia}

\date{\today}

\begin{abstract}
It is shown that variation of the one-particle dispersion in a universal many-body system enables us to obtain Tan's adiabatic sweep theorem and its generalization. The derivation is based on  the Hellmann-Feynman theorem and the variational theorem for the scattering length suggested in our previous paper [Cherny and Shanenko, Phys. Rev. E \textbf{62}, 1646 (2000)]. As an example, the universal effects in the system of spinless bosons are considered. With the help of the variational theorem, we obtain the mean kinetic and interaction energies and derive the virial theorem for the homogeneous and trapped bosons. The results can easily be generalized to the two-component fermions with interactions between opposite spins.
\end{abstract}
\maketitle

\section{Introduction}
\label{sec:intro}
Experimentally, interactions between cold atoms can successfully be controlled with Feshbach resonances \cite{rev:Chin10,Pitaevskii16:book}. However, theoretical description of strongly interacting many-body systems is a challenging task in general. More than ten years ago Tan suggested \cite{Tan2008a,Tan2008b} a simple way of obtaining thermodynamic potentials of strongly interacting cold gases. The suggested thermodynamic relations contain a quantity called contact parameter, which is determined by the long-range tail of the single-particle momentum distribution. Tan's relations for two-component none-polarized Fermi gases were verified both theoretically \cite{Hu11} and experimentally \cite{Stewart10,*Kuhnle11,*Hoinka13}, see Chap.~18 of the textbook \cite{Pitaevskii16:book} and the review \cite{Braaten12}.

Tan's relations are based on the assumption of universality of dilute systems, which implies that their thermodynamic properties depend on interparticle interactions only through the zero-momentum amplitude of the two-body Schr\"odinger equation. The zero-momentum scattering amplitude is proportional to the scattering length. Once the universality is assumed, the real interaction potential can be replaced by the pseudopotential $V(r)=\frac{4\pi\hbar^2a}{m}\delta(\bm{r})$ with the scattering length $a$ as a control parameter. This scheme is very convenient for applying various field-theory methods; however, it might lead to ultraviolet divergencies. Besides, being formally applied, the pseudopotential gives incorrect results, for instance, in finding the kinetic and interaction energies \cite{cherny00,cherny01}.

Tan found \cite{Tan2008b} an increment of the energy when the scattering length changes slowly (adiabatic sweep theorem) or rapidly (dynamic sweep relation). In this paper we consider only the adiabatic effects, which are related to equilibrium thermodynamics.

There are many various derivations of Tan's adiabatic sweep theorem
(see the references in the extensive review \cite{Braaten12}). In this article, we follow the approach going back to Bogoliubov's paper \cite{Bogoliubov47} and developed in our pervious publications \cite{cherny00,cherny01,cherny02,cherny04}. It is to keep the shape of the potential and justify the universality. This approach is free from any divergencies, and we believe that it enhances understanding of the correlation properties of many-body systems.

The main idea of the paper is quite simple, see Sec.~\ref{sec:homsys} below. If a many-body system is universal,
its thermodynamics is determined by the scattering length, which can be considered as a thermodynamic property.
By the Hellmann-Feynman theorem, the single-particle momentum distribution is proportional to the variational derivative with respect to the one-particle dispersion: $\frac{\delta F}{\delta T_{p}} =\left\langle\frac{\delta \hat{H}}{\delta T_{p}}\right\rangle$, where $F$ is the free energy or another corresponding thermodynamic potential. It follows from the universality that $\frac{\delta F}{\delta T_{p}}=\frac{\partial F}{\partial a}\frac{\delta a}{\delta T_{p}}$ near $T_{p}=\frac{\hbar^2p^2}{2m}$. Applying the variational theorem (see Sec.~\ref{sec:scat_leng} below) for the scattering length,
we obtain the relation between the momentum distribution at large momenta and the thermodynamic property $\frac{\partial F}{\partial a}$.
In this way we extend Tan's theorem for interaction potentials of arbitrary shape. The obtained relations (\ref{Tanrelgen}) and (\ref{Tanrelgentrap}) are the main results of the paper.

For the sake of simplicity we consider spinless bosons. For bosons, however, some caveats should be made (see the detailed discussion and references in the review \cite{Chevy16}). The contribution of the Efimovian states at large scattering length
involves an additional three-body parameter
of dimension of length, which makes the system nonuniversal \cite{werner08,werner09,*werner12}. In this paper we consider only universal effects. The obtained results can easily be extended to the two-component fermions with interactions between opposite spins.

The paper is organized in the following way. In the section \ref{sec:scat_leng} we discuss in detail the variational theorem for the scattering length. In the next section, we derive Tan's adiabatic sweep theorem for homogeneous and trapped Bose systems and its extension from the variational theorem. Besides, we find the mean value of kinetic and interaction energies and obtain the virial theorem for the homogeneous and trapped gases. In the Conclusion, the main results of the paper are briefly discussed.

\section{The variational theorem for the scattering length}
\label{sec:scat_leng}

The two-body interaction ${V}(r)$ is supposed to be of the short-range type, that is, it should fall off at infinity as $1/r^{\alpha}$ with the exponent $\alpha>3$ or faster. Note that the exponent is usually equal to six for realistic interatomic potentials (see Ref.~\cite{llvol3_77}, Sec.~89). Let us consider an $s$-wave (that is, radially symmetric) solution of the two-body Schr\"odinger equation in the centre-of-mass system with zero relative momentum
\begin{equation}
-\frac{\hbar^2}{m}\nabla^2\varphi^{}(r)+{V}(r)\varphi^{}(r)=0.
\label{twobody}
\end{equation}
It obeys the boundary conditions: $|\varphi^{}(r)|<\infty$ at $r=0$ and
\begin{equation}
\varphi(r)\simeq 1-a/r
\label{scatasymp}
\end{equation}
when $r\to \infty$. The parameter $a$ is called the scattering length, and it can take arbitrary real value. Since the potential and asymptotics (\ref{scatasymp}) are real then the solution is real as well. It is convenient to separate the scattering part of the wave function $\psi(r)$ defined as
\begin{align}
\varphi^{}(r)=1+\psi^{}(r).
\label{psidef}
\end{align}

In the Fourier representation, the Schr\"odinger equation (\ref{twobody}) reads
\begin{align}
\psi^{}_{p}=-\frac{U(p)}{2T_{p}},
\label{SchFour}
\end{align}
where $T_{p}=\hbar^2 p^2/(2m)$ is the free-particle dispersion, and $U(p)$ is the scattering amplitude "off the mass shell":
\begin{align}
U^{}(p)=&\int{d}^3r\,{V}(r)\varphi^{}(r)e^{i\bm{p}\cdot\bm{r}}\nonumber\\
=&\frac{4\pi}{p}\int_{0}^{\infty}{d}r\,r{V}(r)\varphi^{}(r)\sin pr.
\label{scatamp}
\end{align}

By integrating Eq.~(\ref{twobody}) and using the divergence theorem and asymptotics (\ref{scatasymp}), we arrive at the relation between the scattering length and the scattering amplitude at zero momentum:
\begin{align}
a=&\frac{m}{4\pi\hbar^{2}}U^{}(0),
\label{adef} \\
U^{}(0)=&\int{d}^3r\,\varphi^{}(r){V}(r)=\int _{0}^{\infty}dr\,4\pi r^2\varphi^{}(r){V}(r).
\label{u0bare}
\end{align}
We also write down a useful expression for the scattering amplitude \cite{Bogoliubov47} (see also the discussion in Ref.~\cite{cherny00})
\begin{align}\label{U0rel}
U(0)=-\int _{0}^{\infty}dr\,4\pi r^2\Big[\frac{dV(r)}{d r}r + 2V(r)\Big]\varphi^2(r),
\end{align}
which can be found from the Schr\"odinger equation (\ref{twobody}) and the boundary conditions.

The following variational theorem \cite{cherny00} for the scattering amplitude \cite{note_chshpapers} is used as a basis for proving Tan's adiabatic sweep theorem
\begin{align}
\delta{U}^{}(0)=&\int{d}^3r
\Bigl[\psi^{}(r)\delta\Bigl(-\frac{\hbar^2\nabla^2}{m} \Bigr)\psi^{}(r)
+\delta{V}(r)\varphi^{2}(r) \Bigr]\nonumber
\\
=&\int\frac{{d}^3p}{(2\pi)^3}\,
2\delta T_{p}\psi^{2}(p)+\int{d}^3r\,\delta{V}(r)\varphi^{2}(r).
\label{varth}
\end{align}
In the last equation, the first term is given in the Fourier representation, and the variation is supposed to be taken near $T_{p}=\hbar^2 p^2/(2m)$. We emphasize that the variation $\delta T_{p}$ should obey the condition $\delta T_{p}/T_{p}\to0$ for $p\to0$ in order to leave the asymptotics (\ref{scatasymp}) and relation (\ref{adef}) unchanged. The variation of the scattering amplitude with respect to the interaction potential is known for a long time (see, e.g., Ref.~\cite{Popov79}). The full form of the variational theorem was suggested in Ref.~\cite{cherny00}; it includes the variation with the both one-particle dispersion and interaction potential.

In order to prove this relation, we represent Eq.~(\ref{u0bare}) in the form
\begin{equation}
{U}^{}(0)
=\int{d}^3r\Bigl[\frac{\hbar^2}{m}|\nabla\psi^{}(r)|^2
                            +[\varphi^{}(r)]^2 {V}(r)\Bigr],
\label{33b}
\end{equation}
which can be found from Eqs. (\ref{twobody}) and (\ref{psidef}).
Further, varying  Eq.~(\ref{33b}) and keeping in mind Eqs.~(\ref{twobody}) and (\ref{scatasymp}), we arrive at Eq.~(\ref{varth}).

Finally, we write the variational theorem (\ref{varth}) in the form
\begin{align}
  \frac{\delta U(0)}{\delta T_{p}}
                                   =&\frac{1}{(2\pi)^3}\frac{U^{2}(p)}{2T_{p}^{2}}=\frac{m^2U^{2}(p)}{4\pi^3\hbar^4p^4}, \label{varTp}\\
  \frac{\delta U(0)}{\delta V(r)} =& \varphi^{2}(r).\label{varV}
\end{align}

The short-range potential $V(r)$ is supposed to be localized within the range $r\lesssim r_0$. It follows that first, Eq.~(\ref{scatasymp}) is satisfied when $r\gtrsim r_0$, and second, $U(p)\simeq U(0)$ when $p\lesssim 1/r_0$. If the integral $\int{d}r\, r^4 V(r)$ converges at large distances then one can calculate the correction to the scattering amplitude within this range in terms of the solution of the Schr\"odinger equation (\ref{twobody}). Expanding $\sin p r$ into the Taylor series and substituting it into Eq.~(\ref{scatamp}) yield 
after a little algebra
\begin{align}
U(p)\simeq U(0)\left(1-\frac{p^2}{a}\int_{0}^{\infty}{d}r\, r^2\left[\varphi(r)-1+\frac{a}{r}\right]\right).
\label{u0corr}
\end{align}

In order to get an impression about the generic behaviour of the scattering amplitude $U(p)$, we solve the Sch\"odinger equation (\ref{twobody}) for the square-well potential
\begin{align}\label{vrsq}
 V(r)=\begin{cases}
 -\frac{\hbar^2\varkappa^2}{m r_0^2} ,  &\text{for } r\leqslant r_0,\\
 0, &\text{for } r > r_0.
 \end{cases}
\end{align}
Here $\varkappa$ is a dimensionless positive parameter, controlling the well depth. We obtain the solution $\varphi(r)=\frac{r_0\sin(\varkappa r/r_0)}{r\varkappa\cos\varkappa}$ inside the radius $r_0$, and $\varphi(r)=1-\frac{r_0}{r}\left(1-\frac{\tan \varkappa}{\varkappa}\right)$ outside. The scattering length and the scattering amplitude are given by the formulas
\begin{align}
  a=&\left(1-\frac{\tan \varkappa}{\varkappa}\right)r_0, \label{a_rect}\\
  U(p)=&
\frac{4\pi\hbar^2r_0 \varkappa^2}{m(p^2r_0^2-\varkappa^2)}\left(\cos p r_0\frac{\tan \varkappa}{\varkappa}
  -\frac{\sin pr_0}{pr_0}\right),\label{Up_rect}
\end{align}
respectively. The expansion of $U(p)$ at small momenta is obtained directly from the last equation or from the general relation (\ref{u0corr}):
\begin{align}\label{Upexp}
  \frac{U(p)}{U(0)}=1 + p^2r_0^2 \left[\frac{1}{\varkappa^2}-\frac{1}{2} + \frac{\varkappa}{3 (\varkappa - \tan \varkappa)}\right]+\cdots.
\end{align}

\begin{figure}[!tbp]
\begin{center}
\includegraphics[width=.8\columnwidth,clip=true]{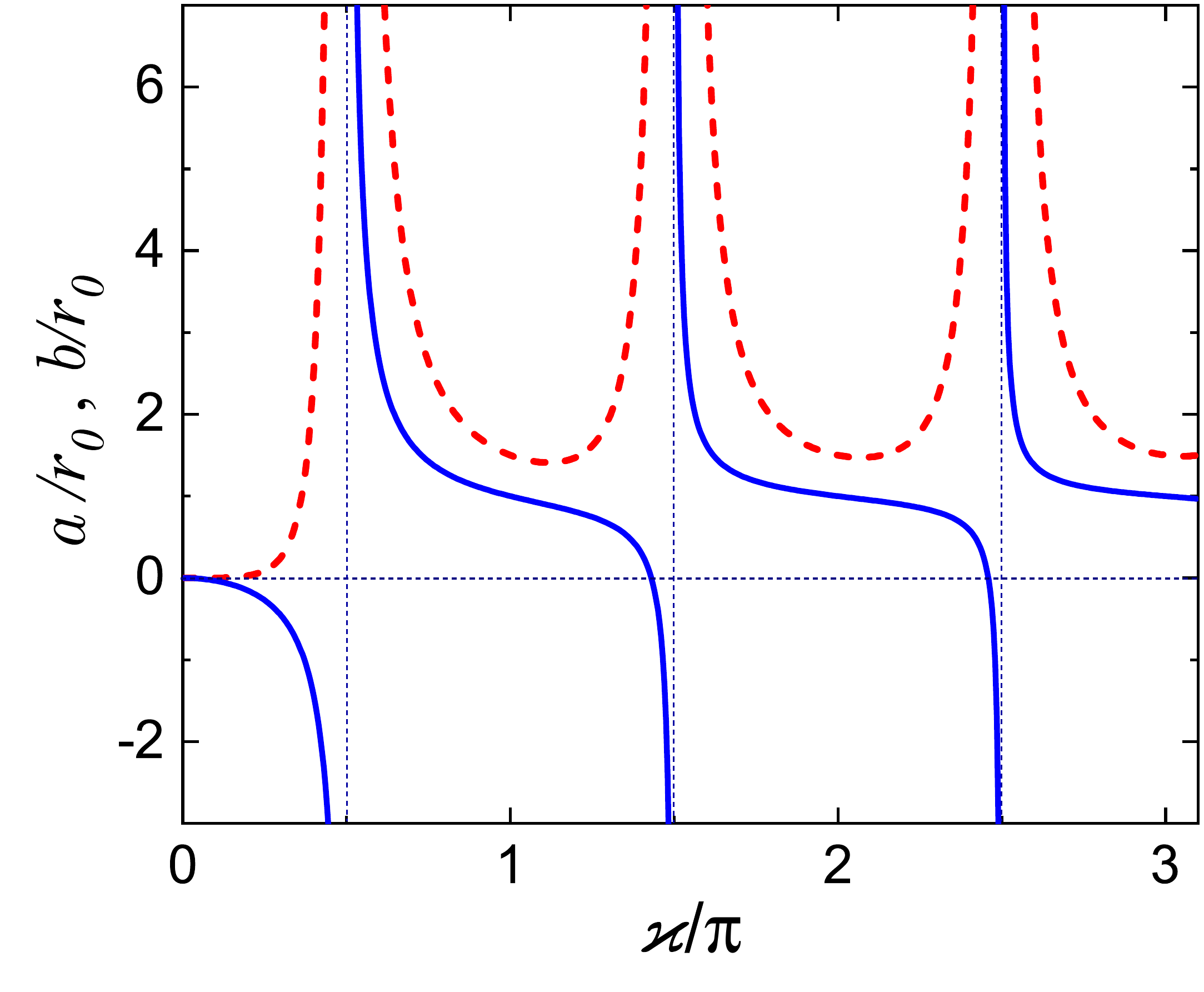}
\end{center}
\caption{\label{fig:ab} The scattering length $a$ (\ref{a_rect}) (blue solid line) and the parameter $b$  (\ref{bdef}) (dashed red line) for the square-well potential (\ref{vrsq}) as a function of the dimensionless control parameter $\varkappa$ (in units of $\pi$). The parameters $a$ and $b$ are given in units of the range of the potential $r_0$. The resonances correspond to the appearing bound states at $\varkappa=\frac{\pi}{2}(1+2 k)$ with $k=0,1,\cdots$. The parameter $b$ is always non-negative.
}
\end{figure}
\begin{figure}[!tbp]
\begin{center}
\includegraphics[width=.9\columnwidth,clip=true]{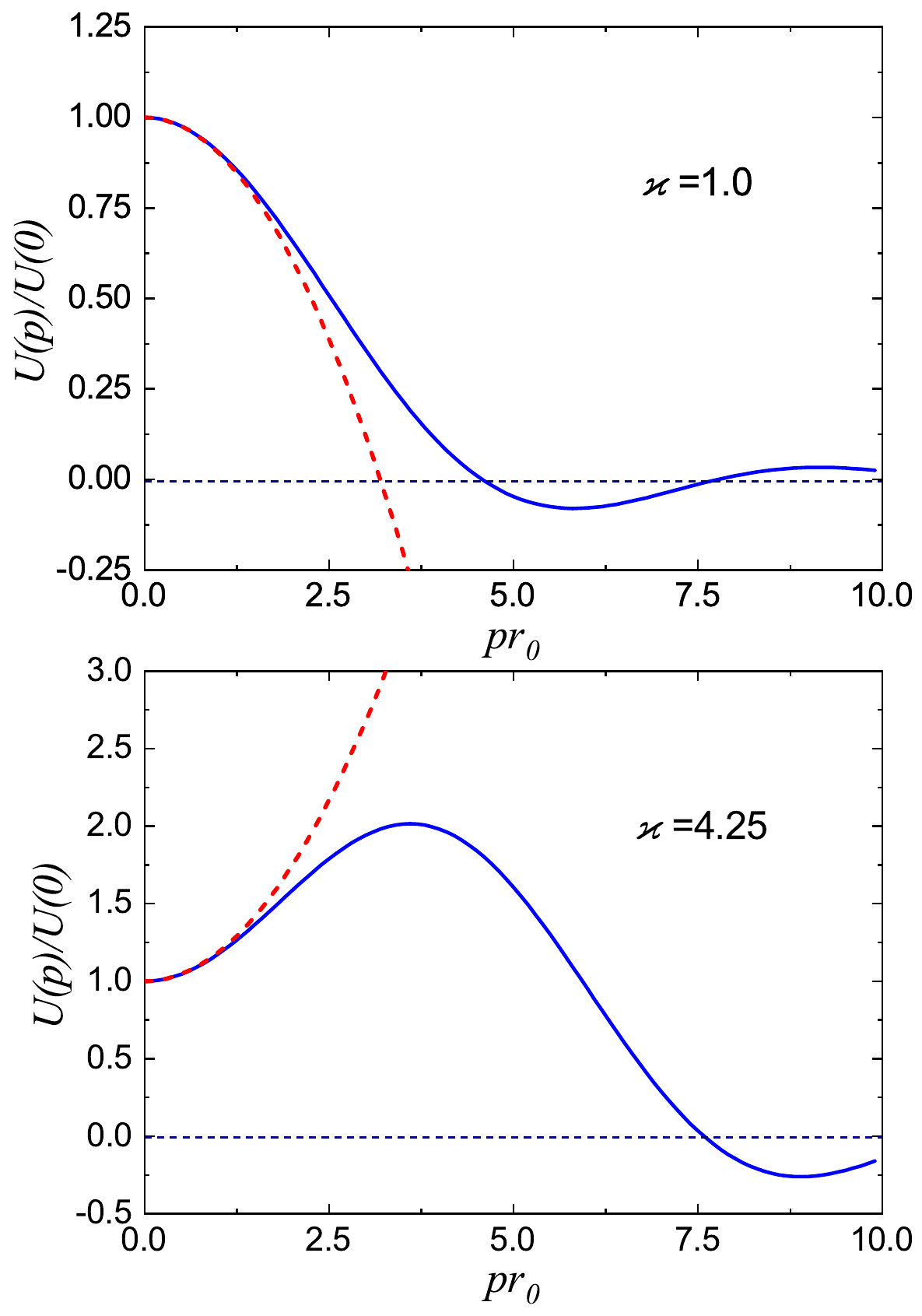}\\
\end{center}
\caption{\label{fig:Up} The scattering amplitude (\ref{scatamp}) (in units of $U(0)={4\pi\hbar^2 a}/{m}$) for the square-well potential (\ref{vrsq}) as a function of the wavenumber (in units of the inverse potential radius $1/r_0$) at different values of the control parameter $\varkappa$. The exact values of the amplitude, given by Eq.~(\ref{Up_rect}), are shown in solid blue line, while its expansion (\ref{Upexp}) at small momenta is represented in dashed red line. The approximate formula (\ref{u0corr}) works well for $p r_0\lesssim 1$.
}
\end{figure}

The scattering length as a function of the potential depth exhibits a set of resonances at $\varkappa=\frac{\pi}{2}(1+2 k)$ with $k=0,1,\cdots$, see Fig.~\ref{fig:ab}.  The behaviour of the scattering amplitude is shown in Fig.~\ref{fig:Up}. A small momenta, the amplitude can increase or decrease, and, hence, the expansion coefficient in Eq.~(\ref{Upexp}) can be positive or negative, respectively. At large wave vectors $p\gg1/r_{0}$, $U(p)$ tends to zero, which is its generic behaviour.

\section{Tan adiabatic sweep theorem and other relations}
\label{sec:Tanrel}

\subsection{Homogeneous systems}
\label{sec:homsys}

Let us consider a system of interacting spinless bosons. The Hamiltonian of the system takes the general form
\begin{align}
\hat{H}=& -\frac{\hbar^2}{2m}\!\int\! d^3r{\hat\Psi}^{\dag}(\bm{r})\nabla^2{\hat\Psi}(\bm{r})\!+\!\int\! d^3 r\,V_\mathrm{ext}(\bm{r}){\hat\Psi}^\dag(\bm{r}){\hat\Psi}(\bm{r})\nonumber\\
&+\frac{1}{2}\!\int\! d^3r_1 d^3r_2 V(|\bm{r}_1\!-\!\bm{r}_2|) {\hat\Psi}^{\dag}(\bm{r}_1){\hat\Psi}^{\dag}(\bm{r}_{2}){\hat\Psi}(\bm{r}_{2}){\hat\Psi}(\bm{r}_1),
\label{Ham}
\end{align}
where ${\hat\Psi}(\bm{r})$ and ${\hat\Psi}^{\dag}(\bm{r})$ are the bosonic field operators obeying the commutation relations $[{\hat\Psi}(\bm{r}),{\hat\Psi}^{\dag}(\bm{r}')]=\delta(\bm{r}-\bm{r}')$. Here $V_\mathrm{ext}(\bm{r})$ is an external potential; for instance, it can be a harmonic trapping potential.

The kinetic energy term in the Hamiltonian (\ref{Ham}) reads in the momentum representation
\begin{align}
\hat{H}_\mathrm{kin}=\sum_{\bm{p}} T_{p}\hat{n}_{\bm{p}}\simeq\frac{V}{(2\pi)^3}\int{d}^3p\, T_{p}\hat{n}_{\bm{p}},
\label{Hamkin}
\end{align}
where the second equality is the main asymptotics in the thermodynamic limit, and $V$ is the volume. Here $\hat{n}_{\bm{p}}=\hat{a}^{\dag}_{\bm{p}}\hat{a}_{\bm{p}}$ is the occupation number operator of wave-vector state $\bm{p}$, and $\hat{a}^{\dag}_{\bm{p}}$ and $\hat{a}_{\bm{p}}$ are the creation and annihilation operators, respectively. They are related to the field operators by ${\hat\Psi}(\bm{r}) =\sum_{\bm{p}}\hat{a}_{\bm{p}}\frac{e^{i\bm{p}\cdot\bm{r}}}{\sqrt{V}}$ and ${\hat\Psi}^{\dag}(\bm{r}) =\sum_{\bm{p}}\hat{a}_{\bm{p}}^{\dag}\frac{e^{-i\bm{p}\cdot\bm{r}}}{\sqrt{V}}$.
Their commutator is calculated as $[\hat{a}_{\bm{p}},\hat{a}^{\dag}_{\bm{p}'}]=\delta_{\bm{p},\bm{p}'}$.  If the interactions are radially symmetric, the average occupation numbers are symmetric as well: $\langle\hat{n}_{\bm{p}}\rangle=n_{p}$.

The main assumption is the universality of the system under consideration: at sufficiently low densities the particles "feel" each other only through the scattering length. Therefore, the energy of the system or another thermodynamic potential depends on interactions through the scattering length $a$. Experimentally, the scattering length can be controlled with Feshbach resonance \cite{rev:Chin10}, which leads to the idea to treat the length $a$ as an intensive thermodynamic property. This means that in the canonical ensemble, we have for the free energy $F=F(T,V,N,a)$ and
\begin{align}\label{dE}
dF=-SdT-pdV+\mu dN +x_{a}da,
\end{align}
where we denote $x_{a}=\left(\partial F/\partial a\right)_{T,V,N}$. Thus we introduce a new pair of conjugated thermodynamic properties, $a$ and $x_{a}$. Once we know the equation of state $x_{a}=x_{a}(T,V,N,a)$, we can integrate Eq.~(\ref{dE}) along a path in the space of thermodynamic variables $(T,V,N,a)$ starting from $a=0$ or $a=\infty$ and thus calculate the free energy for arbitrary $a$.

However, this scheme remains useless until we do not specify a method of measuring $x_a$. Tan's brilliant idea is to relate this thermodynamic property with the momentum distribution of particles at high momenta, which is experimentally accessible. This is the essence of Tan's adiabatic sweep theorem.

Let us prove Tan's theorem. The Hellmann--Feynman theorem tells us that
a variation of free energy with respect to parameters of the Hamiltonian
is given by $\delta F =\langle\delta \hat{H}\rangle$.
Here the brackets stand for the average over the canonical ensemble: $\langle\cdots\rangle=\tr(\cdots e^{-\beta\hat{H}})/\tr e^{-\beta\hat{H}}$ and $\beta=1/T$ (we put the Boltzmann constant equal to one).
The Hellmann-Feynman theorem is also applicable in the microcanonical and grand canonical ensemble with the corresponding averages: $\delta E =\langle\delta \hat{H}\rangle$ and $\delta \Omega =\langle\delta \hat{H}\rangle$, respectively. Here $E$ is the mean energy, and $\Omega$ is the grand potential.

First we find the momentum distribution at high momenta. Let us vary the one-particle dispersion $T_{p}$ near $T_{p}=\hbar^2p^2/(2m)$ while keeping the mass constant. It follows from Eqs.~ (\ref{Ham}) and (\ref{Hamkin}) and the Hellmann-Feynman theorem that variation with respect to the one-particle dispersion yields the average occupation numbers $\frac{\delta F}{\delta T_{p}} =\left\langle\frac{\delta \hat{H}}{\delta T_{p}}\right\rangle =V\frac{n_{p}}{(2\pi)^3}$. On the other hand, at constant mass, the free-energy variation depends on $\delta T_{p}$ only through the variation of the scattering length: $\frac{\delta F}{\delta T_{p}}=\frac{\partial F}{\partial a}\frac{\delta a}{\delta T_{p}}$. Equating these expressions and using Eq.~(\ref{adef}) and the variation theorem (\ref{varTp}), we arrive at the relation
\begin{align}
p^4 n_{p}=\frac{1}{V}\frac{8\pi m}{\hbar^2}\left(\frac{\partial F}{\partial a}\right)_{T,V,N}a^2\frac{U^{2}(p)}{U^{2}(0)},
\label{Tanrelgen}
\end{align}
where $U(p)$ is the scattering amplitude (\ref{scatamp}).

The obtained relation (\ref{Tanrelgen}) generalizes Tan's adiabatic sweep theorem. It is valid for arbitrary $p\gtrsim 1/\xi$, where $\xi$ is a parameter with the dimension of length, which determines the characteristic scale of the many-body effects. In the case of the dilute Bose gas, $\xi= 1/\sqrt{4\pi a n}$ is the healing length, provided  $na^3\ll 1$, where $n=N/V$ is the density of particles. For large $na^3$, the parameter $\xi$ is of order of mean distance between particles: $\xi\sim n^{-1/3}$ with the caveats discussed in Sec.~\ref{sec:intro} above. Within the range $1/\xi \lesssim p\lesssim 1/r_0$ ($r_0$ is the radius of the interaction potential), the scattering amplitude $U(p)$ is almost constant: $U(p)\simeq U(0)=4\pi \hbar^2 a/m$, see the discussion in Sec.~\ref{sec:scat_leng} above.
Then we arrive at Tan's theorem \cite{Tan2008b}
\begin{align}\label{Tanrel}
p^4 n_{p}\simeq{\cal C}=\frac{1}{V}\frac{8\pi m}{\hbar^2}\left(\frac{\partial F}{\partial a}\right)_{T,V,N} a^2,
\end{align}
where ${\cal C}$ is the parameter introduced by Tan \cite{Tan2008a} and called Tan's contact. Note that for the pseudopotential, we have formally $U(p)=4\pi \hbar^2 a/m=\text{const}$ and $r_0=0$, and, therefore, Tan's relation (\ref{Tanrel}) is obtained from Eq.~(\ref{Tanrelgen}) as a particular case.

We emphasize that for realistic interactions, the limit $\lim_{p\to\infty} p^4 n_{p}$ is actually equal to zero, since by Eq.~(\ref{Tanrelgen}) $p^4 n_{p}$ is proportional to the squared scattering amplitude, which tends to zero at large momenta (see Fig.~\ref{fig:Up}). However, for cold atoms, the values of the wave-vector of order $1/r_0$ are hardly accessible experimentally, and one can talk about "experimentally reasonable" limit, taking a non-zero  value. Nevertheless, the generalized relation (\ref{Tanrelgen}) can be considered as a \emph{fitting} formula for the thermodynamic property ${\partial F}/{\partial a}$, which takes into account the corrections to the scattering amplitude $U(0)$. In this case, we should introduce a model interatomic interaction potential, solve the Schr\"odinger equation (\ref{twobody}), and use the exact equation (\ref{scatamp}) or its approximation (\ref{u0corr}).

The short-range spatial correlations are obtained in the same manner \cite{cherny00,cherny01}. The Hellmann-Feynman theorem yields
$\frac{\delta F}{\delta V(r)} =\left\langle\frac{\delta \hat{H}}{\delta V(r)}\right\rangle =\frac{ N(N-1)}{2V} g(r)$
where
$g(r) =g(\bm{r}) =\frac{V^2}{N^(N-1)} \langle{\hat\Psi}^{\dag}(\bm{r}){\hat\Psi}^{\dag}(0){\hat\Psi}(0){\hat\Psi}(\bm{r})\rangle$
is the pair distribution function, which is proportional to the density-density correlator. Here we use the invariance of correlation functions under a constant shift of the coordinates in the absence of external forces.
The function $g(\bm{r})$ admits a clear physical interpretation as the conditional probability density of finding a particle at the point $\bm{r}$ provided another particle is located at the origin of coordinates.
On the other hand, we have from Eqs.~ (\ref{adef}) and (\ref{varV}) that $\frac{\delta F}{\delta V(r)} =\frac{\partial F}{\partial a}\frac{\delta a}{\delta V(r)} =\frac{\partial F}{\partial a}\frac{m}{4\pi\hbar^2}\varphi^2(r)$. Finally, we obtain in the thermodynamic limit
\begin{align}
g(r)=&w\,\varphi^2(r), \label{varchSh}\\
w=&\frac{1}{V}\left(\frac{\partial F}{\partial a}\right)_{T,V,N}\frac{m}{2\pi n^2\hbar^2}.
\label{varVr}
\end{align}
Besides, the factor $w$ can be written in terms of Tan's contact (\ref{Tanrel})
\begin{align}\label{varVrTan}
w=\frac{{\cal C}}{16\pi^2a^2n^2}.
\end{align}
The equation (\ref{varchSh}) is valid at short distances $r\lesssim \xi$. It was obtained for the dilute Bose gas at zero temperature by the same method twenty years ago \cite{cherny00,cherny01}. As was shown \cite{cherny00}, the density expansion of the factor (\ref{varVr}) is $w= w(na^{3})= 1 +\frac{64}{3\sqrt{\pi}}\sqrt{na^3} +\cdots$, and the result (\ref{varchSh}) is in excellent agreement with the Monte Carlo simulations (see the discussion in Ref.~\cite{cherny01}).  The relation (\ref{varchSh}) together with (\ref{varVrTan}) generalizes the equation for the pair correlation function, obtained within the pseudopotential approach \cite{Tan2008a}
\begin{align}
 g(r)\simeq w\,\left(1-\frac{a}{r}\right)^2,\label{grps}
\end{align}
which is applicable only when $r_0 \lesssim r\lesssim \xi$.

\subsection{Virial theorem for a homogeneous system}
\label{sec:virial}
The short-range behaviour (\ref{varchSh}) of the pair correlation function is found from the assumption that thermodynamics is determined by the scattering length. Reversely, Eq.~(\ref{varchSh})  can be used for obtaining various thermodynamic relations. Here we derive the relation between pressure and energy per volume for a system of spinless bosons.

The virial theorem for a homogeneous system of spinless particles (see, e.g., Ref.~\cite{book:zubarev74}) gives us the following expression for the pressure
\begin{equation}
P=\frac{2}{3}\frac{\langle\hat{H}_\mathrm{kin}\rangle}{V} -\frac{n^{2}}{6} \int_{0}^{\infty}dr\,4\pi r^2\frac{dV(r)}{d r}r g(r),
\label{virialth}
\end{equation}
where $\langle\hat{H}_\mathrm{kin}\rangle$ is the mean value of the kinetic energy operator (\ref{Hamkin}). On the other hand, the mean energy per volume in the thermodynamic limit is given by
\begin{align}
\frac{E}{V}=\frac{\langle\hat{H}_\mathrm{kin}\rangle}{V} +\frac{n^2}{2}\int_{0}^{\infty}dr\,4\pi r^2 V(r)g(r),
\label{enbos1}
\end{align}
Eliminating the kinetic energy term from Eqs.~(\ref{virialth}) and (\ref{enbos1}) yields
\begin{equation}
P=\frac{2}{3}\frac{E}{V} -\frac{n^{2}}{6} \int_{0}^{\infty}dr\,4\pi r^2 \left[\frac{dV(r)}{d r}r +2V(r)\right]g(r).
\label{pressure}
\end{equation}
After substituting the pair distribution function given by Eqs.~(\ref{varchSh}) and (\ref{varVrTan}) into Eq.~(\ref{pressure}) and using the formula (\ref{U0rel}), we are left with the equation
\begin{equation}
P=\frac{2}{3}\frac{E}{V}+\frac{\hbar^2{\cal C}}{24\pi a m}.
\label{pressurefin}
\end{equation}

As one can see, the intricate integral in Eq.~(\ref{pressure}) is elegantly and compactly transformed into a function of the scattering length, as it should be. This method goes back to the original Bogoliubov  paper \cite{Bogoliubov47}, and it was developed in our previous publications \cite{cherny00,cherny04}. The form of this equation through Tan's contact was suggested in Ref.~\cite{Tan2008c} for the non-polarized two-component Fermi gas with interactions between opposite spins. In this case, the last term differs by a factor of two.

\subsection{Kinetic and interaction energies}
\label{sec:kinint}

Even when thermodynamics of a system is successfully described by the scattering length, not all equilibrium quantities can be written in terms of it. The pseudopotential approach fails here, because it contains only one parameter --- the scattering length. A good example is the kinetic and interaction energies of the three-dimensional dilute Bose gas \cite{cherny00,cherny02,cherny04} (see also the discussion in Sec.~4.3 of Ref.~\cite{Pitaevskii16:book}).

\subsubsection{Interaction energy}
\label{sec:int}

The interaction energy per volume can be calculated with the already obtained pair distribution function given by Eqs.~(\ref{varchSh}) and (\ref{varVrTan}):
\begin{align}
  \frac{E_\mathrm{int}}{V}=&\frac{n^2}{2}\int{d}^3r\, V(r)g(r)=\frac{\hbar^2{\cal C}}{8\pi m}\frac{(a-b)}{a^2},\label{int}\\
  b=&\frac{1}{4\pi}\int d^{3}r\,\bigl|\nabla\varphi(r)\bigr|^{2}=\int_{0}^{\infty}dr\, r^2\left[\frac{\partial \varphi(r)}{\partial r}\right]^{2}.\label{bdef}
\end{align}
Here we use Eqs.~(\ref{adef}) and (\ref{33b}) and introduce the characteristic length $b$, which is always positive \cite{cherny00}. The integral in Eq.~(\ref{bdef}) converges, since $\frac{\partial \varphi(r)}{\partial r }\sim {1}/{r^2}$ at large distances. Note one subtlety here: the expression (\ref{varchSh}) for the pair distribution function is valid only at the distances $r\lesssim \xi$, while the short-range interacting potential $V(r)$ is localized within $r\lesssim r_0\ll \xi$. Then the long-range tail of $g(r)$ is not significant in calculating the interaction energy.

The behaviour of the parameter $b$ for the square-well potential (\ref{vrsq}) is shown in Fig.~\ref{fig:ab}. For more realistic interatomic potentials, containing a strongly repulsive core and attractive tail proportional to $1/r^{6}$, the the parameter $b$ is by order of magnitude larger than the scattering length $a$ \cite{cherny02}.

Another way \cite{cherny00} to find the interaction energy is to replace $V(r)\to \lambda V(r)$, vary $\lambda$ near $\lambda=1$ and apply the Hellmann-Feynman theorem. It follows from the variational theorem (\ref{varth}) that
\begin{align}
\left.\frac{\partial a}{\partial \lambda}\right|_{\lambda=1}=m\frac{\partial a}{\partial m}=a-b.
\label{amlambda}
\end{align}
Again, the parameter $\lambda$ appears in the free energy only though the scattering length, and hence $E_\mathrm{int} =\left.\frac{\partial F}{\partial \lambda}\right|_{\lambda=1} =\left.\frac{\partial F}{\partial a}\frac{\partial a}{\partial \lambda}\right|_{\lambda=1}$. Then we arrive at Eq.~(\ref{int}) with the help of Eqs.~(\ref{Tanrel}) and (\ref{amlambda}).

\subsubsection{Kinetic energy}
\label{sec:kin}

The mean kinetic energy cannot be obtained directly by integrating the one-particle dispersion $T_{p}$ multiplied by the mean occupation number $n_{p}$ from Eq.~(\ref{Tanrelgen}), because
this equation is valid only for large momenta, while the contribution of small momenta to the kinetic energy is significant. However, we can vary the free energy with respect to the mass of particles, keeping in mind that it depends on the mass not only through the scattering length. From dimensionless considerations \cite{Pitaevskii16:book}, the free energy
can be written down as
\begin{align}\label{frendim}
F(T,V,N,a)=N\frac{\hbar^2n^{2/3}}{m}\tilde{f}\left(na^3,\frac{Tm}{\hbar^2n^{2/3}}\right),
\end{align}
where $\tilde{f}$
is a function of two dimensionless variables. With the Hellmann-Feynman theorem, we find the mean kinetic energy as
${E_\mathrm{kin}}
=- m\frac{\partial F}{\partial m}=F -T\left(\frac{\partial F}{\partial T}\right)_{V,N,a}-\left(\frac{\partial F}{\partial a}\right)_{V,N,a}m\frac{\partial a}{\partial m}$. The first two terms are equal to the mean energy, because $E=F+TS$, and the last term coincides with the interaction energy due to Eq.~(\ref{amlambda}). Finally, we are left with the equation
\begin{align}\label{Ekin}
{E_\mathrm{kin}} =F -T\left(\frac{\partial F}{\partial T}\right)_{V,N,a}-{E_\mathrm{int}}=E-{E_\mathrm{int}},
\end{align}
where $E_\mathrm{int}$ is given by Eq.~(\ref{int}). The interaction energy is determined by Tan's contact, while the kinetic energy is not, and  the total energy is the sum of the both. Then the developed scheme is fully consistent.

\subsection{Inhomogeneous systems}
\label{sec:inhomsys}

We consider the system of bosons in a radially symmetric harmonic trap. The Hamiltonian is given by Eq.~(\ref{Ham}) with $V_\mathrm{ext}(r)=k r^2/2$, where $k=m\omega^2$ is the stiffness coefficient of the harmonic forces. The kinetic energy term in the Hamiltonian (\ref{Ham}) can be written as
\begin{align}
\hat{H}_\mathrm{kin}=
\frac{1}{(2\pi)^3}\int{d}^3p\, T_{p}\hat{\Psi}^{\dag}_{\bm{p}}\hat{\Psi}_{\bm{p}},
\label{Hamkintrap}
\end{align}
where ${\hat\Psi}(\bm{r}) =\frac{1}{(2\pi)^{3}}\int{d}^3p\,\hat{\Psi}_{\bm{p}}e^{i\bm{p}\cdot\bm{r}}$ and ${\hat\Psi}^{\dag}(\bm{r}) =\frac{1}{(2\pi)^{3}}\int{d}^3p\,\hat{\Psi}^{\dag}_{\bm{p}}e^{-i\bm{p}\cdot\bm{r}}$. The operators
obey commutation relations $[\hat{\Psi}_{\bm{p}},\hat{\Psi}^{\dag}_{\bm{p'}}] =(2\pi)^3 \delta(\bm{p}-\bm{p'})$ with $\delta(\bm{p})$ being the $\delta$-function. Then the momentum distribution is given by $N(p) =N(\bm{p}) =\langle\hat{\Psi}^{\dag}_{\bm{p}}\hat{\Psi}_{\bm{p}}\rangle$; it is normalized to the total number of particles: $\frac{1}{(2\pi)^{3}}\int {d}^3 p\,N(\bm{p}) =N$.  The distribution $N(p)$ has the dimension of volume, and it is analogous to $n_{p}V$ considered in Sec.~\ref{sec:homsys} above.

From the dimension considerations, the free energy takes the form
\begin{align}\label{frendimtrap}
F(T,\omega,N,a)=\hbar\omega\tilde{f}\left(\frac{Na}{a_\mathrm{ho}},\frac{T}{\hbar\omega}\right),
\end{align}
where $a_\mathrm{ho}=\hbar/\sqrt{m\omega}$ is the oscillator length. Tan's adiabatic sweep theorem is proved by analogy with the homogeneous case by means of the Hellmann-Feynman theorem: $\frac{N(p)}{(2\pi)^3} =\left\langle\frac{\delta \hat{H}}{\delta T_{p}}\right\rangle=\frac{\delta F}{\delta T_{p}}=\frac{\partial F}{\partial a}\frac{\delta a}{\delta T_{p}}$. With the variational theorem (\ref{varTp}), we arrive at the relations, analogous to Eqs.~(\ref{Tanrelgen}) and (\ref{Tanrel}), respectively,
\begin{align}
p^4 N({p}) =&\,\frac{8\pi m}{\hbar^2}\left(\frac{\partial F}{\partial a}\right)_{T,\omega,N}a^2\frac{U^{2}(p)}{U^{2}(0)},
\label{Tanrelgentrap}\\
p^4 N({p})\simeq&\,{\cal I}=\frac{8\pi m}{\hbar^2}\left(\frac{\partial F}{\partial a}\right)_{T,\omega,N} a^2,\label{Tanreltrap}
\end{align}
where ${\cal I}$ is Tan's constant for trapped gases.
The limits of validity of Eqs.~(\ref{Tanrelgentrap}) and (\ref{Tanreltrap}) are the same as that of Eqs.~(\ref{Tanrelgen}) and (\ref{Tanrel}), respectively. However, the many-body characteristic length $\xi$ is inhomogeneous and depend on the local density of particles; it should be taken at the center of the trap, where the density is highest.

The short-range spatial correlations are calculated in the same manner: $\frac{\delta F}{\delta V(r)} =\left\langle\frac{\delta \hat{H}}{\delta V(r)}\right\rangle = \frac{1}{2}\int d^3\!R\, n^{2}(R)g(R,r)$, where $g(R,r)= { \langle{\hat\Psi}^{\dag}\left(\bm{R}+\frac{\bm{r}}{2}\right){\hat\Psi}^{\dag}\left(\bm{R}-\frac{\bm{r}}{2}\right){\hat\Psi}\left(\bm{R}-\frac{\bm{r}}{2}\right)
{\hat\Psi}\left(\bm{R}+\frac{\bm{r}}{2}\right)\rangle}/{n^{2}(R)}$ is the local pair distribution function and $n(R) =\langle{\hat\Psi}^{\dag}(\bm{R}){\hat\Psi}(\bm{R})\rangle$ is the local density of particles. On the other hand, we have $\frac{\delta F}{\delta V(r)} =\frac{\partial F}{\partial a}\frac{\delta a}{\delta V(r)} =\frac{\partial F}{\partial a}\frac{m}{4\pi\hbar^2}\varphi^2(r)$ from the variational theorem. Then we obtain from Eq.~(\ref{Tanreltrap})
\begin{align}\label{gRrI}
\int d^3\!R\, n^{2}(R)g(R,r)=\frac{{\cal I}}{16\pi^2a^2}\varphi^2(r).
\end{align}

This equation gives us the pair distribution function, averaged over the squared density profile. The characteristic scales of $g(R,r)$ with respect to $R$  and $r$ are $a_\mathrm{ho}$ and $\xi$, respectively. They usually differ by orders of magnitude, which enables us to apply the local density approximation $g(R,r)=g(n(R),r)$ with $g(n,r)$ being the pair distribution function found in the homogeneous case. Then Eqs.~(\ref{varchSh}) and (\ref{varVrTan}) yield $n^{2}(R)g(R,r)=\frac{{\cal C}(n(R))}{16\pi^2 a^2}\varphi^2(r)$. Substituting this expression into  Eq.~(\ref{gRrI}), we arrive at the relation between ${\cal I}$ and ${\cal C}$ within the local density approximation
\begin{align}\label{ICrel}
 {\cal I}=\int d^3\!R\, {\cal C}(n(R)).
\end{align}

\subsection{Virial theorem, kinetic and interaction energies in the inhomogeneous case}
\label{sec:virkinint}

In the same manner, we calculate the mean oscillator energy: $E_\mathrm{ho} =\left\langle k\frac{\partial \hat{H}}{\partial k}\right\rangle =k\frac{\partial F}{\partial k}$. Using the relations $k\frac{\partial \omega}{\partial k}=\frac{\omega}{2}$ and $k\frac{\partial a_\mathrm{ho}}{\partial k}=-\frac{a_\mathrm{ho}}{4}$, we find from Eq.~(\ref{frendimtrap}) after a little algebra
\begin{align}
 E_\mathrm{ho} =\frac{1}{2} \left[F-T\left(\frac{\partial F}{\partial T}\right)_{\omega,N,a}\right] +\frac{a}{4}\left(\frac{\partial F}{\partial a}\right)_{T,\omega,N},\nonumber
\end{align}
and, hence,
\begin{align}
E_\mathrm{ho} =\frac{E}{2}+\frac{\hbar^2{\cal I}}{32\pi m a},
\label{Eho}
\end{align}
where Eq.~(\ref{Tanreltrap}) is used. The mean full energy is given by the sum $E=E_\mathrm{kin}+E_\mathrm{int}+E_\mathrm{ho}$, and we obtain the virial theorem \cite{werner08}
\begin{align}\label{virth}
 2E_\mathrm{rel}-2E_\mathrm{ho} = -\frac{\hbar^2{\cal I}}{8\pi m a},
\end{align}
where we introduce the release energy $E_\mathrm{rel}=E_\mathrm{kin}+E_\mathrm{int}$. Note that the virial theorem (\ref{pressurefin}) in the homogeneous case can be obtained by the same method from the free energy (\ref{frendim}).

The interaction and kinetic energies can also be found with the Hellmann-Feynman theorem by analogy with the homogeneous case (see Sec.~\ref{sec:kinint}): $E_\mathrm{int} =\left.\frac{\partial F}{\partial \lambda}\right|_{\lambda=1} =\left.\frac{\partial F}{\partial a}\frac{\partial a}{\partial \lambda}\right|_{\lambda=1}$ and ${E_\mathrm{kin}}=- m\frac{\partial F}{\partial m}$. Equations (\ref{amlambda}), (\ref{frendimtrap}), and  (\ref{Tanreltrap}) lead to
\begin{align}\label{Einttrap}
E_\mathrm{int} =\frac{\hbar^2{\cal I}}{8\pi m}\frac{a-b}{a^2}.
\end{align}
The kinetic energy is obtained the same way with the help of the relations $m\frac{\partial \omega}{\partial m}=-\frac{\omega}{2}$ and $m\frac{\partial a_\mathrm{ho}}{\partial m}=-\frac{a_\mathrm{ho}}{4}$:
\begin{align}\label{Ekintrap}
E_\mathrm{kin} =\frac{E}{2}-\frac{\hbar^2{\cal I}}{32\pi m a}-\frac{\hbar^2{\cal I}}{8\pi m}\frac{a-b}{a^2}.
\end{align}
The sum of oscillator, interaction, and kinetic energies is equal to the full energy, as it should be.

\section{Conclusion}

The obtained equations for the homogeneous (\ref{Tanrelgen}) and trapped (\ref{Tanrelgentrap}) systems of bosons are applicable for wave vectors more than $\xi$, which is of the order of the mean distance between particles. The equations are valid even for $p\gtrsim 1/r_{0}$, where $r_{0}$ is the radius of the interaction potential. They generalize Tan's adiabatic sweep theorem and can be used as fitting formulas for the thermodynamic property ${\partial F}/{\partial a}$, which takes into consideration the prefactor $U^2(p)/U^2(0)$. The analogous equations can be written down for the non-polarized two-component fermions.

Once the derivative ${\partial F}/{\partial a}$ is known, the pair distribution function given by Eqs.~(\ref{varchSh}) and (\ref{varVr}) is also known at distances $r\lesssim\xi$. It generalizes the usual expression (\ref{grps}) obtained within the pseudopotential approach. The prospect of finding the structure factor within the range of wave vectors $p\gtrsim 1/\xi$ is quite interesting. Note that the structure factor obtained from Eq.~(\ref{grps}) by means of the Fourier transformation is restricted to the range $1/\xi \lesssim p\lesssim 1/r_{0}$ \cite{Pitaevskii16:book,Braaten12}.


The variational theorem (\ref{varth}) is also useful for obtaining the mean kinetic and interaction energies for the homogeneous (Sec.~\ref{sec:kinint}) and inhomogeneous (Sec.~\ref{sec:virkinint}) many-body systems.

\section{Acknowledgement}

The author acknowledges support from the JINR--IFIN-HH projects.

\bibliography{tanvar}

\begin{thebibliography}{24}
\bibitem [{not()}]{note_chshpapers}%
  \BibitemOpen
  \href@noop {} {}\bibinfo {note} {In the papers \cite{cherny00,cherny01}, the
  solution of the two-body Schr\"odinger equation and the scattering amplitude
  are denoted as $\varphi^{(0)}(r)$ and $U^{(0)}(p)$,
  respectively.}\BibitemShut {Stop}%
\end{thebibliography}

\end{document}